# Computer Aided Design Modeling for Heterogeneous Objects

Vikas Gupta[1], K.S.Kasana[2], and Puneet Tandon[3]

[1]Research Scholar, NIT Kurukshetra, India
Haryana, India

[2]National Institute of Technology
Kurukshetra, India

[3] PDPM-IIITDM,
Jabalpur, India

**Abstract**
Heterogeneous object design is an active research area in recent years. The conventional CAD modeling approaches only provide geometry and topology of the object, but do not contain any information with regard to the materials of the object and so can not be used for the fabrication of heterogeneous objects (HO) through rapid prototyping. Current research focuses on computer-aided design issues in heterogeneous object design. A new CAD modeling approach is proposed to integrate the material information into geometric regions thus model the material distributions in the heterogeneous object. The gradient references are used to represent the complex geometry heterogeneous objects which have simultaneous geometry intricacies and accurate material distributions. The gradient references helps in flexible manipulability and control to heterogeneous objects, which guarantees the local control over gradient regions of developed heterogeneous objects. A systematic approach on data flow, processing, computer visualization, and slicing of heterogeneous objects for rapid prototyping is also presented.
*Keywords: HO, Gradient Reference, Visualization, Slicing, Rapid Prototyping.*

## 1. Introduction

The recent research focuses on computer-aided design issues involved in rapid prototyping of heterogeneous object. The primary goal of the present research is to develop systematic methodologies for heterogeneous object representations, visualizations, constructions and manipulations. The term 'heterogeneous object' is defined such that it can have different material composition within an object. There are three subclasses of heterogeneous object e.g. [12]:

- Multiple materials object.
- Object with sub-objects embedded.
- Object without clear material boundary (Functionally graded materials, FGM).

Traditional CAD systems, used for conventional design method, can only represent the geometry and topology of an object. No material information is available within the representation which is required for heterogeneous objects. With the capability to fabricate heterogeneous objects, functionally efficient and cost reducing designs can be realized. Rapid prototyping (RP) techniques allow heterogeneous material objects to be produced using 3D CAD models by varying material composition region-wise, layer-wise, or point-wise. The required 3D CAD model should have not only the geometric information but also the information of materials, property, etc. at each point inside an object. In order to take full advantage of the greatest potential of heterogeneous objects, one must have matching capabilities for their computer modeling, analysis, design optimization and visualization. The primary focus of the recent research development in these fields is on the computer representation schemes for heterogeneous objects, by extending the mathematical models and computer data structures of the modern solid modeling techniques to include discrete material regions of interfacial boundaries and heterogeneous properties. Recent studies show that an effective heterogeneous CAD modeling system should at least meet the following specifications e.g. [4]:

- Intuitive in representing geometry, topology and material information simultaneously.
- Capable of representing complex solids: the solids to be modeled may be complex in geometry as well as in material variations.
- Compact and exact: the representation should be compact, and both the geometry and material

- information can be retrieved accurately and efficiently.
- The representation of material properties must be compatible with current or proposed standards for geometric modeling representations as described in ISO 10033. This is essential to exchange data among design, analysis and manufacturing process plan domains.

This paper is organized as follows: in section 2, the previous work is reviewed; Section 3 is brief representation of procedure for CAD data flow and processing of HO; Section 4 represents developed mathematical model and address different gradient references for the local control of gradient regions; heterogeneous object visualization related issues are studied in section 5; slicing procedure for the fabrication of heterogeneous objects is discussed in section 6 and the final section is conclusion with future scope to extend this work.

## 2. Review of Research Work

Approaches of modeling of HO have been extensively studied in computer and manufacturing community. Kumar and Dutta proposed an approach to model multi-material objects based on R-m sets and R-m classes primarily for application in layered manufacturing. Boolean operators were defined to facilitate the modeling process e.g. [5-6]. Jackson et al. proposed a local composition control (LCC) approach to represent heterogeneous object in which a mesh model is divided into tetrahedrons and different material compositions are evaluated on the nodes of the tetrahedrons by using Bernstein polynomials e.g. [3],[7]. Chiu developed material tree structure to store different compositions of an object e.g. [2]. The material tree was then added to a data file to construct a modified format being suitable for RP manufacturing. Siu and Tan developed a scheme named 'source-based' method to distribute material primitives, which can vary any material with an object e.g. [12]. The feature-based modeling scheme was extended to heterogeneous object representation through boundary conditions of a virtual diffusion problem in the solid, and then designers could use it to control the material distribution e.g. [10-11]. Liu extended his work in by taking parameterized functions in terms of distance(s) and functions using Laplace equation to smoothly blend various boundary conditions, through which designers could edit geometry and composition simultaneously [4],[10]. Kou and Tan suggested a hierarchical representation for heterogeneous object modeling by using B-rep to represent geometry and a heterogeneous feature tree to express the material distributions e.g. [4]. Various methods for designing and optimizing objects composed of multiple regions with continuously varying material properties have been developed. Wang and Wang proposed a level-set based variational scheme [12]. Biswas et al. presented a mesh-free approach based on the generalized Taylor series expansion of a distance field to model and analyze a heterogeneous object satisfying the prescribed material conditions on a finite collection of material features and global constraints [13-14]. However, almost all of the research interests are mainly focused on the computer representation of heterogeneous object, rather than the whole procedure for rapid prototyping fabrication of heterogeneous object. The approaches were verified in commercial software packages, such as Solidworks and Unigraphics [7],[10]. A commercial CAD package independent system is developed to deal with the HO modeling, but not including the slicing procedure for RP manufacturing [11]. In this paper, we just address the CAD gradient reference model with systematic methodologies for visualization and manipulation of heterogeneous objects.

## 3. CAD Data Flow and Processing of HO

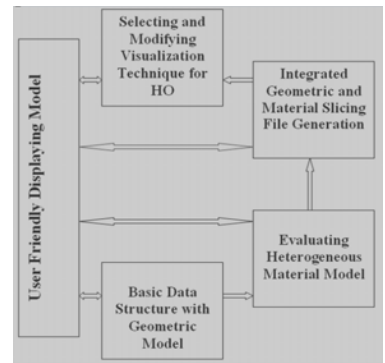

Fig. 1 CAD data flow for the development of HO.

The structure of CAD data flow system of HO contains five main modules;(1) basic data structure module;(2) evaluating heterogeneous material information model;(3) HO dataset visualization module;(4) Slice generation with gradient material information; and (5) display module e.g. figure (1).

The data processing module mainly copes with the data structure set up for geometric model and the subdivision surfaces for improving the smoothness of meshes. The second module evaluates material information of a gradient region within a CAD model according to the specifications of the users. In our system, we exploit the geometric model to describe the shape information. In terms of material information, it describes material composition in terms of material space. Third module mainly provides visualization and rendering information for visualizing heterogeneous objects. Rapid prototyping technique offers a possibility to manufacture heterogeneous object. The accuracy and quality of the final part fabricated by rapid prototyping depends on the 2D geometric slices of a model. Fourth module gives the slice generating information by describing the material information in a layer. Fifth module displays the complete information on user friendly display.

The RP processes are dependent on a CAD model of the heterogeneous object which generates the required information for driving the RP machine. The necessary tasks to generate this information are termed as process planning tasks. RP processes can fabricate heterogeneous objects by selectively depositing various materials in a point-wise fashion using 3D-CAD data representation without special tooling. In these processes, a uniform layer of powder is spread over the built area and the different layers are joined together by different methods to form the prototype. The information flow for processing the heterogeneous describes the necessity of developing material modeling system along with the geometric model e.g. figure (2).

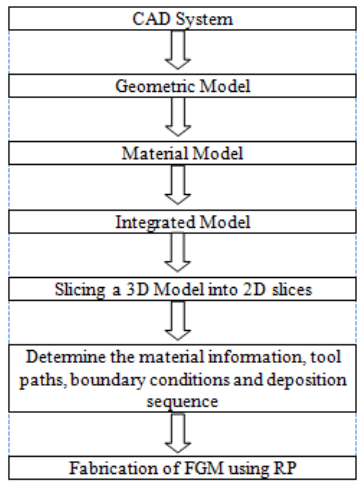

Fig. 2 Processing of the HO.

## 4. Mathematical Representation of HO

The proposed model represents intricate geometries as well as material variation simultaneously; assures smooth material variations throughout the complex object and local material alterations. A noteworthy point to be emphasized here is that in the proposed model, the topological information is utilized to ensure smooth material variations throughout the complex geometry heterogeneous object. In current work, mathematically the heterogeneous object is defined as:

$$P = (G, M) = (P_1, P_2, \ldots P_n) \quad (1)$$
$$P_i = (G_i, M_c) \quad (2)$$
$$\sum_{i=1}^{n} G_i = P \text{ and } G_i = \sum_{j=1}^{m} G_{ij} \quad (3)$$

Where P is the heterogeneous object with geometric information G and material information M. P is also a set of n number of cells, where $P_i$ represents $i^{th}$ cell in the object with the geometric information ($G_i$), occupied by a m number of sub volumes and specific material distribution for each cell ($M_c$). $M_c$ represents material composition of pre-defined number of primary materials in $P_i^{th}$ cell with $G_i^{th}$ geometric information. The accuracy of the model is increased by having local control on sub volumes ($G_{ij}$) in a cell at various identified locations which also results in less huge storage space problem.

### 4.1 Material Composition Function

Material composition function f(s) is a function of distance from the end point of first homogeneous region to the first differential geometric point (where the material gradient becomes zero). Either linear or non-linear analytical function that fall on real domain can be use, with the distance from point to the grading reference as variable, therefore it is exact, e.g. the material distribution function with FGM at a distance 'a' & up to a distance '(1-a)' is described e.g. Eq. (4).

$$f(s) = \begin{cases} 0, & s \leq a \\ f(s), & a < s < (1-a) \\ 1, & s \geq (1-a) \end{cases} \quad (4)$$

The effect of logarithmic and power functions on material distribution in a gradient region can be visualized e.g. figure (3).

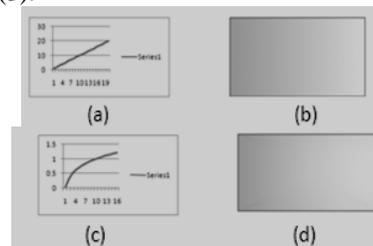

Fig.3 Material distribution for different functions: (a) power function, (b) material distribution for power function, (c) logarithmic function, (d) material distribution for logarithmic distribution. which also results in less huge storage space problem.

### 4.2 Material Composition Array

Each element of the material composition array $M_c$ represents the volume fraction of pre-defined primary materials in $G_i^{th}$ cell. The total volume fraction of the primary materials for the material composition arrays should be summed up to one. The end material composition arrays of heterogeneous region are denoted by **M$_{cs}$** and **M$_{cf}$**, for start and final boundary of heterogeneous region respectively. Generally, if k materials are included in the object, then:

$$\sum_{r=1}^{k} M_{csr} = 1 \text{ and } \sum_{r=1}^{k} M_{cfr} = 1 \qquad (5)$$

Where, $M_{csr}$ = r$^{th}$ material element of the material composition $M_{cs}$
$M_{cfr}$ = r$^{th}$ material element of the material composition array $M_{cf}$.
k = number of primary materials including air

So, material composition of any primary material $V_j$ for a sub volume $G_{ij}$ is find out e.g. Eq.(6).

$$V_j = f(s) \times (M_{csr} - M_{cfr}) + M_{csr}, \begin{cases} M_{csr} \in M_{cs} \\ M_{cfr} \in M_{cf} \\ 0 \leq f(s) \leq 1 \end{cases} \qquad (6)$$

The property of heterogeneous unit volume is determined using Voigt's rule e.g. Eq. (7).

$$S = \sum_{j=1}^{m} \sum_{r=1}^{k} V_j S_r \qquad (7)$$

S is the property of heterogeneous volume fraction
$V_j$ is the volume fraction of each material in unit volume
$S_r$ is the property of r$^{th}$ material.
For two materials composition, the heterogeneous property is defined e.g. Eq. (8).

$$S = V_1 S_1 + (1 - V_1) S_2 \qquad (8)$$

Sub-volume creation algorithm results in sharp material changes along the component boundaries, which potentially result in abrupt property (e.g. thermal expansion coefficient and stiffness) variations. So for the smooth material transition properties, one of the available blending functions may be used. In our case, the constant blending functions are incorporated at end positions and in between, a blending function, $f_b$, along with the distance function is used to avoid the sharp change in material properties e.g. figure (4).

$$S = f(s)V_1 S_1 + (1 - V_1)(1 - f(s)) f_b S_2 \qquad (9)$$

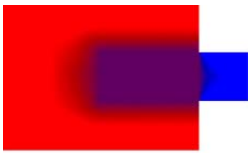

Fig. 4 Heterogeneous object with blended function.

### 4.3 Gradient Reference and Control

The gradient references are start and end boundaries of gradient region which may be controlled locally and provide the required information about gradient origins and respective material distributions. The grading references may be classified into three categories:
- Basic gradient references.
- Offset gradient references.
- Hybrid gradient references.

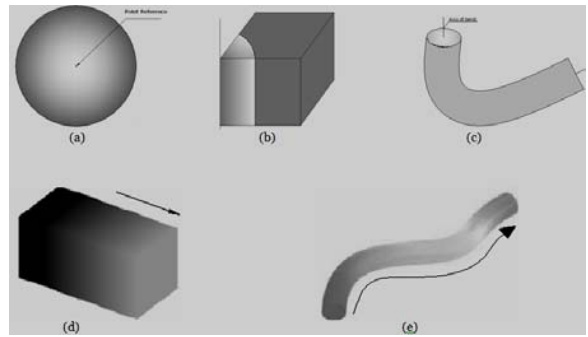

Fig.5 Basic gradient references: (a) point gradient reference, (b) linear axis gradient references, (c) flexible axis gradient reference, (d) linear plane gradient references, and (e) flexible axis plane gradient reference.

The basic gradient references includes user defined point, line and plane or may be any 3D object entity i.e. axis, vertex, edge, axis, any surface of the object. The material distribution for these basic grading references is shown in figure (5). The flexibility in axis and plane gradient references is provided by introducing local control using sweep operation.

The region with offset grading references is divided into a number of sub-regions; further sub-division may be required in order to represent the complex gradient in the local area as shown in figure (6). This can be done by applying recursive sub-division algorithm to either boundary of each sub-region and material composition array (M$_c$) for i$^{th}$ sub-region with r$^{th}$ contour (S$_{br}$) is derived e.g. Eq. (10).

$$M_c(S_{br}, i) = \frac{i \times (M(C_r) - M(C_{r+1}))}{r_m} + M(C_{r+1}) \quad \text{for inwards sub} - \text{division}$$

$$= \frac{(r_m - i) \times (M(C_r) - M(C_{r+1}))}{r_m} + M(C_{r+1}) \quad \text{for outwards sub} - \text{division}$$

$$(10)$$

Where r represents the r$^{th}$ contour of total r$_m$ contours, i = 1,2,………….r-1, and M(C$_r$) represents material composition associated with r$^{th}$ contour.

Specifically, this algorithm can be applied to any of two adjacent contours. However, for sub-regions due to the consideration of material continuity and homogeneity on the surface, only boundaries should be chosen as the geometrical offset reference. The step width array (w) of the composition change in sub-region can be calculated e.g. Eq. (11).

$$w(S_{br}) = M_c(S_{br}, i+1) - M_c(S_{br}, i) \qquad (11)$$

$w(S_{br})$ is a constant vector in the same 2D sub-region since a linear interpolation algorithm is applied between two adjacent contours. In addition to this step width array, the other information needed for fabrication of each sub-region is the material composition array corresponding to either of the two adjacent contours and the number of sub-regions. The computer memory is greatly saved as linear composition gradient within each sub-region is adopted. With the proposed recursive material evaluation algorithm, the material composition of any point inside the object can be exactly evaluated at runtime.

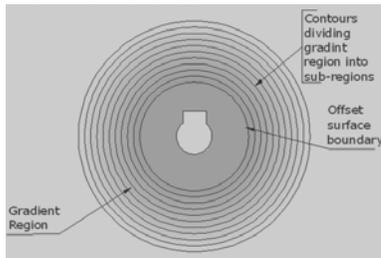

Fig.6  Offset gradient reference with sub-division of region.

The object with dissimilar boundary closures results in requirement of hybrid gradient references. The region with hybrid references is divided into a number of sub-regions as described above in case of offset gradient references e.g. figure (7(a)). However distance blending function with linear interpolation smoothing function is used to develop continuous gradient regions. Blending functions for the object region formed by a local $C^1$ smooth transition between two or more primary surfaces, which may or may not intersect, are used. These blending functions allow the construction of constant radius blends for any type of surfaces as long as their offset surfaces are smooth, without singularities and self intersections. Edge blends are created by sweeping rational quadratic curves. Corner blends are created by a convex combination of Taylor interpolants. The resultant material distribution in hybrid region HO is shown figure (7(b)).

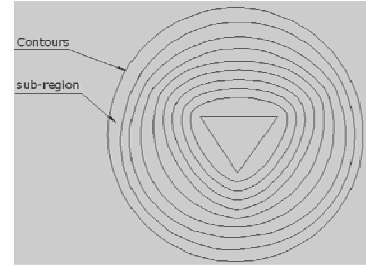

(a)

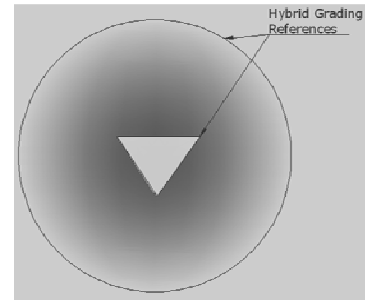

(b)

Fig.7 Hybrid gradient references: (a) sub-division of hybrid reference region, (b) material distribution of hybrid reference HO.

The proposed heterogeneous object model has local and universal control over the gradient references. The current model is an unevaluated representation, which is independent of universal co-ordinate system. The effect of grading on the properties of heterogeneous objects can be easily modified by controlling the respective gradient references. Moreover different material composition functions i.e. linear, exponential, parabolic, power or any other type of functions are used for different grading effects. The effect of changing the gradient references only is shown in figure (8(a)) and figure (8(b)) and the effect of changing the gradient references with the application of recursive sub-division algorithm, distance blending function and smoothing function using linear interpolation is also shown figure 8(c), figure 8(d). The material distribution is remained linear e.g. figure 8(b)., while adapt the shape of changed gradient reference as shown in figure 8(d).

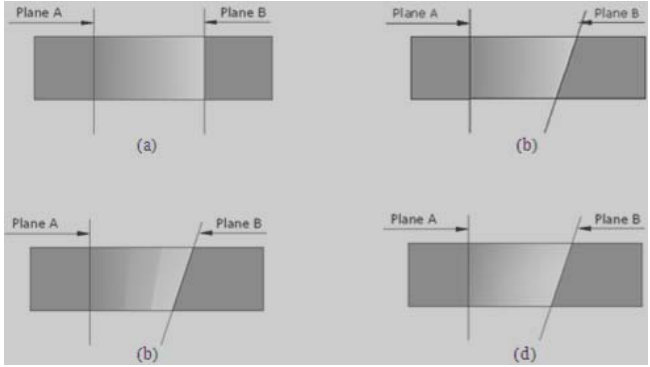

Fig. 8 Local control: (a) linear material distribution between parallel gradient reference planes, (b) effect of changing the gradient references only i.e. linear material distribution exists, (c) sub-division of gradient region using distance blending function, and (d) smoothing of sub-regions using linear interpolation functions.

## 5. Data Processing and Visualization of HO

It includes two main sub-modules called data processing and visualization. The data processing module mainly copes with the data structure set up for geometric model and the subdivision surfaces for improving the smoothness of meshes if a rough mesh model is the input object. In this case, a recursive algorithm is used to subdivide the surfaces while maintaining the sharp features of the object. Triangular meshes are used for sub-division of sub-regions of all geometric models. The geometric and material model are integrated to describe the shape and material information in a three dimensional HO space. At the end the dataset output files are displayed on the windows graphical user interface using many input-output functions and open graphical languages.

Effective and efficient visualization of heterogeneous objects is important to computer-aided design of heterogeneous objects. In the past few decades, visualization of homogeneous solids has undergone extensive studies and a variety of 3D visualization toolkits have been available. In recent years, great attentions have also been paid to volume renderings, which attempt to represent the entire 3D data in 2D images.

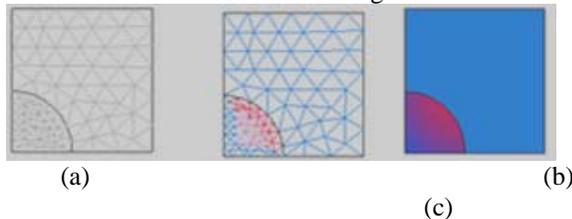

(a)                (b)
(c)

Fig.9: Visualization of HO: (a) sub-division of sub-region using triangular mesh, (b) material evaluation and color mapping of mesh, (c) filling of color to obtain HO gradient region.

However, most of these volume visualization schemes are targeted for rendering volume data obtained by samplings techniques (e.g. MRI or CT images); general issues on efficient visualization of heterogeneous (homogeneous, multi-material, functionally graded material) objects have not been thoroughly studied before. Two major considerations, the visualization efficiency and rendering fidelity, are considered while finalizing the visualization for heterogeneous objects.

Boundary visualization scheme, e.g. figure 9(a), is used to render the external and internal parts of heterogeneous objects. Boundary visualization is intuitive to convey both the shapes and the material distributions of the objects. To render the shapes, the exact boundaries of the objects are first faceted into discrete elements, which are generally termed as boundary meshes. The material distributions within each element are then rendered by using colors or grey values to represent the material compositions. All the rendered facets provide an approximate polygonal view of the heterogeneous objects, in both geometries and material distributions. Efforts to generate faithful, high quality computer visualizations at interactive rate are presented to avoid abrupt material transition effects. The boundary mesh generation is extensively used in traditional solid modeler for homogeneous object renderings. Each face of the object is faceted into triangle meshes and these triangles are then transferred to rendering engines (OpenGL) to generate graphical outputs. Boundary sub-faceting approaches are used; to speed up the visualization process, adaptive boundary sub-faceting scheme and repetitive computation eliminations are introduced. All these approaches show that the proposed visualization scheme can generate effective visualizations in interactive heterogeneous object design.

Once the boundary meshes of the object are obtained, the material compositions of the mesh nodes are then evaluated by using the proposed recursive material evaluation algorithm. The evaluated material composition at a given location is represented with a k-dimensional vector $[r_1, r_2, …, r_k]$, whose element $r_i$ represents the material volume fraction of the $i^{th}$ predefined primary material ( k is the total number of predefined primary materials). A color mapping which maps the material composition to a system color is shown in figure (9(b)). The most commonly used color mapping techniques include RGB (red-green-blue) color mapping and the HLS (hue-lightness-saturation) mapping. Finally, respective colors are filled to obtain heterogeneous objects e.g. figure (9(c)). All these approaches show that the proposed visualization scheme can generate effective visualizations in interactive heterogeneous object design.

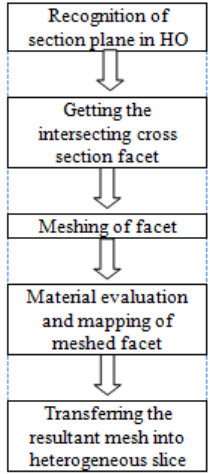

Fig.10 Procedure of sliced generation of the HO.

## 6. Slice Generation of Heterogeneous Objects

Rapid prototyping technique offers a possibility to manufacture HO. The accuracy and quality of the final part fabricated by rapid prototyping depends on the 2D geometric slices of a model. The slices of the geometric model and the layers of the material dataset can be used to construct the 2D slices of heterogeneous object, which is called material resample with geometric constraint. The slicing algorithms are studied extensively in rapid prototyping community. There are mesh-based, direct, adaptive and hybrid slicing algorithms. In our framework, a mesh model slicing algorithm is developed. Sliced generation procedure is proposed to display the internal structures and material distributions in each region of the HO e.g. figure (10) and (11).

The heterogeneous object is first intersected with a section plane and the intersection curves are obtained e.g. figure (11(b)). Then 2D region is faceted (sub-faceted when needed) into meshed region as shown e.g. figure (11(c)). Material evaluation and mapping of meshed facet are sequentially applied e.g. figure (11(d)). The evaluated meshes are then transferred to the heterogeneous slice e.g. figure (11(e)).

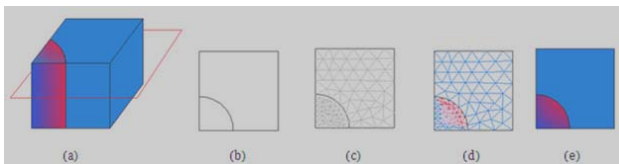

Fig.11: Slice generation procedure: (a) HO with intersection plane, (b) 2D extracted region, (c) meshed region, (d) color mapping against material information, (e) slice of HO. bottom of the page and column where it is cited. Footnotes should be rare.

## 7. Conclusion and Future Scope

This work presents a CAD modeling approach for heterogeneous object with complex geometry and simple material variations. The sequence of CAD data flow for HO and procedure for processing of HO for rapid prototyping processes are discussed. The distribution of material is obtained by using different grading references with local control. Data processing and visualization methods for heterogeneous objects are proposed. Slice generation methodology, necessary required for rapid prototyping of HO is evaluated. The proposed CAD modeling approach represents intricate geometries as well as material variation simultaneously; ensures smooth material variations throughout the complex object; imposes only local material alterations on the cells so that their original properties can be properly retained in the resultant object; offers flexible material variation; consistent in data representations; and computationally robust and efficient.

The present work can be further extended and implemented complex and irregular material distributions. The approach can be extended to object modeling i.e. solid modeling with other physical attributes such as mechanical properties, material distribution etc. Dynamic heterogeneous objects (DHO) are the new class of heterogeneous objects. Unlike current heterogeneous object modeling, DHO deals with space dependent heterogeneities and time dependent shapes and material distributions. By taking time into consideration, more realistic process simulation can be achieved. DHO technology has emerging applications in life science domain, biomedical applications, dynamic process simulation and bio-CAD etc.

**Gupta Vikas,** BE-1999, ME-2003, 10 yrs experience in Technical Education, Currently doing research work as research scholar under Ph.D. program of NIT Kurukshetra, India. Life Membership of ISTE, India. No. of published papers-seven.  Rapid Prototyping, Geometric Modeling, Robotics.

**Kasana K. S.,** Ph.D., Professor, Mechanical Engineering Department, National Institute of Technology, Kurukshetra, India. Guided a number of Ph.D. and M.Tech. thesis.

**Tandon Puneet,** Ph.D., Professor, Mechanical Engineering Department, PDPM-IIITDM, Jabalpur, India. Guided a number of Ph.D. and M.Tech. thesis.